\documentclass[3p]{elsarticle}

\usepackage{hyperref}
\usepackage{amssymb}
\usepackage{bm}
\usepackage{dcolumn}
\usepackage{amssymb}
\usepackage{amsmath}
\usepackage{graphicx}
\usepackage{booktabs}
\usepackage{multirow}
\usepackage{setspace}
\usepackage{array}
\usepackage{threeparttable}
\usepackage{txfonts}
\usepackage{subfig}

\journal{PLB, published in Phys. Lett. B 760 (2016) 602-604.}

\begin{document}

\begin{frontmatter}



\newcommand{\e}{{\mathrm{e}}}
\renewcommand{\i}{{\mathrm{i}}}
\renewcommand{\deg}{^\circ}

\newcommand*{\PKU}{School of Physics and State Key Laboratory of Nuclear Physics and
Technology, Peking University, Beijing 100871,
China}
\newcommand*{\CIC}{Collaborative Innovation Center of Quantum Matter, Beijing, China}
\newcommand*{\CHEP}{Center for High Energy Physics, Peking University, Beijing 100871, China}
\newcommand*{\CHPS}{Center for History and Philosophy of Science, Peking University, Beijing 100871,
China}

\title{Light speed variation from gamma ray burst GRB~160509A}

\author[a]{Haowei Xu}
\author[a,b,c,d]{Bo-Qiang Ma\corref{cor1}}

\address[a]{\PKU}
\address[b]{\CIC}
\address[c]{\CHEP}
\address[d]{\CHPS}
\cortext[cor1]{Corresponding author \ead{mabq@pku.edu.cn}} 

\begin{abstract}
It is postulated in Einstein's relativity that the speed of light in vacuum is a constant for all observers. However, the effect of quantum gravity could bring an energy dependence of light speed. Even a tiny speed variation, when amplified by the cosmological distance, may be revealed by the observed time lags between photons with different energies from astrophysical sources. From the newly detected long gamma ray burst GRB~160509A, we find evidence to support the prediction for a linear form modification of light speed in cosmological space.
\end{abstract}


\end{frontmatter}

At $T_0$=08:59:07.16 UT on 9 May 2016, the Gamma-ray Burst Monitor (GBM)~\cite{GBM}~onboard the Fermi Gamma-ray Space Telescope  (FGST) was triggered by photon flux from the gamma ray burst GRB~160509A~\cite{GBM_trigger} (hereafter all times are measured relative to $T_0$). About one day later, beginning at 13:15 UT on 10 May 2016, the Gemini North Telescope on Mauna Kea observed the optical counterpart of GRB~160509A and revealed a single, well detected emission line which is interpreted as $\mathrm{[OII]}~3727~AA$ at a redshift of $z=1.17$~\cite{160509Az}.

\begin{figure}[!h]
\centering
\subfloat[]{
\label{fig:1a}
\begin{minipage}{0.4\textwidth}
\centering
\includegraphics[width=\textwidth]{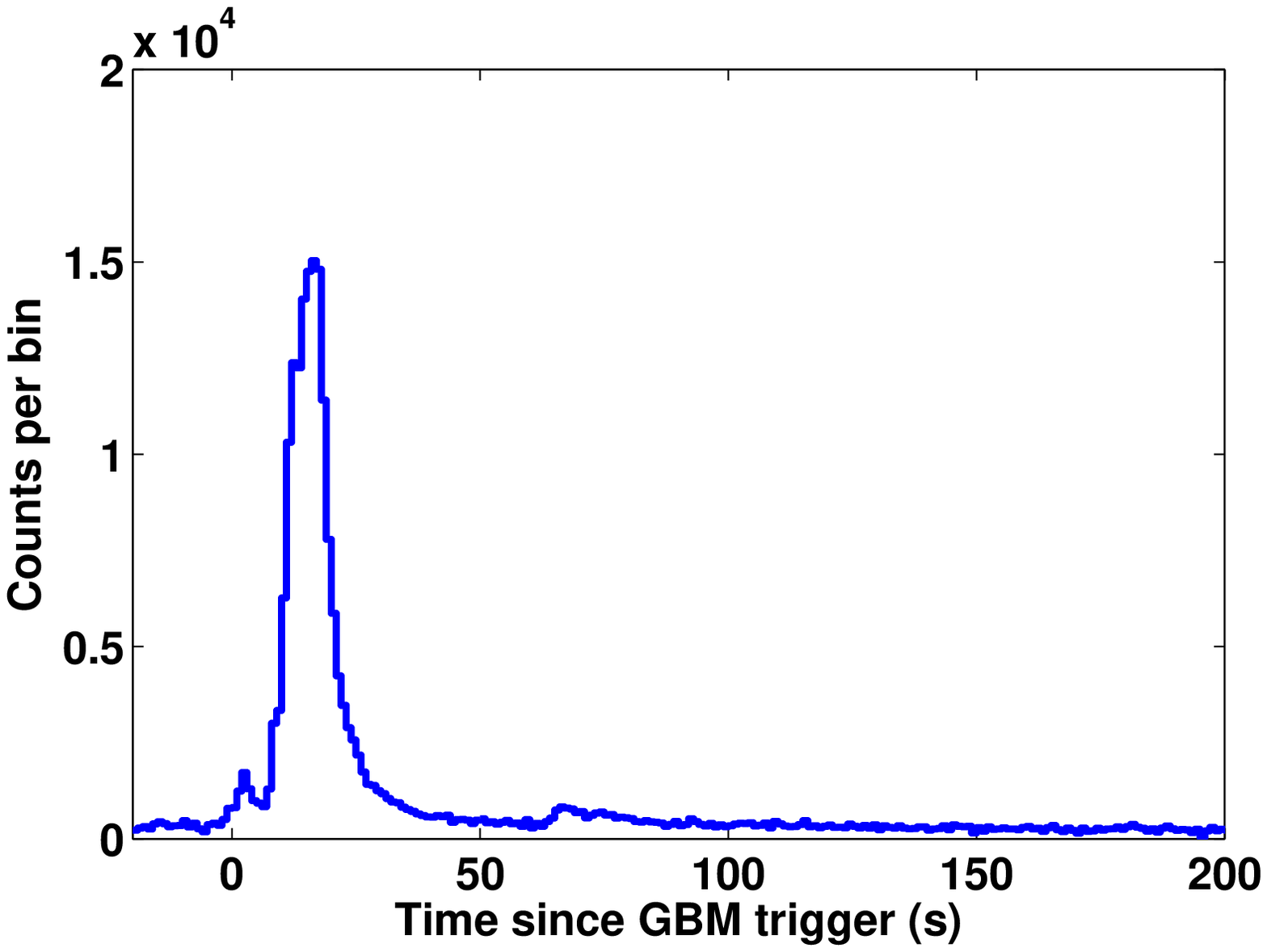}
\end{minipage}
}
\subfloat[]{
\label{fig:1b}
\begin{minipage}{0.4\textwidth}
\centering
\includegraphics[width=\textwidth]{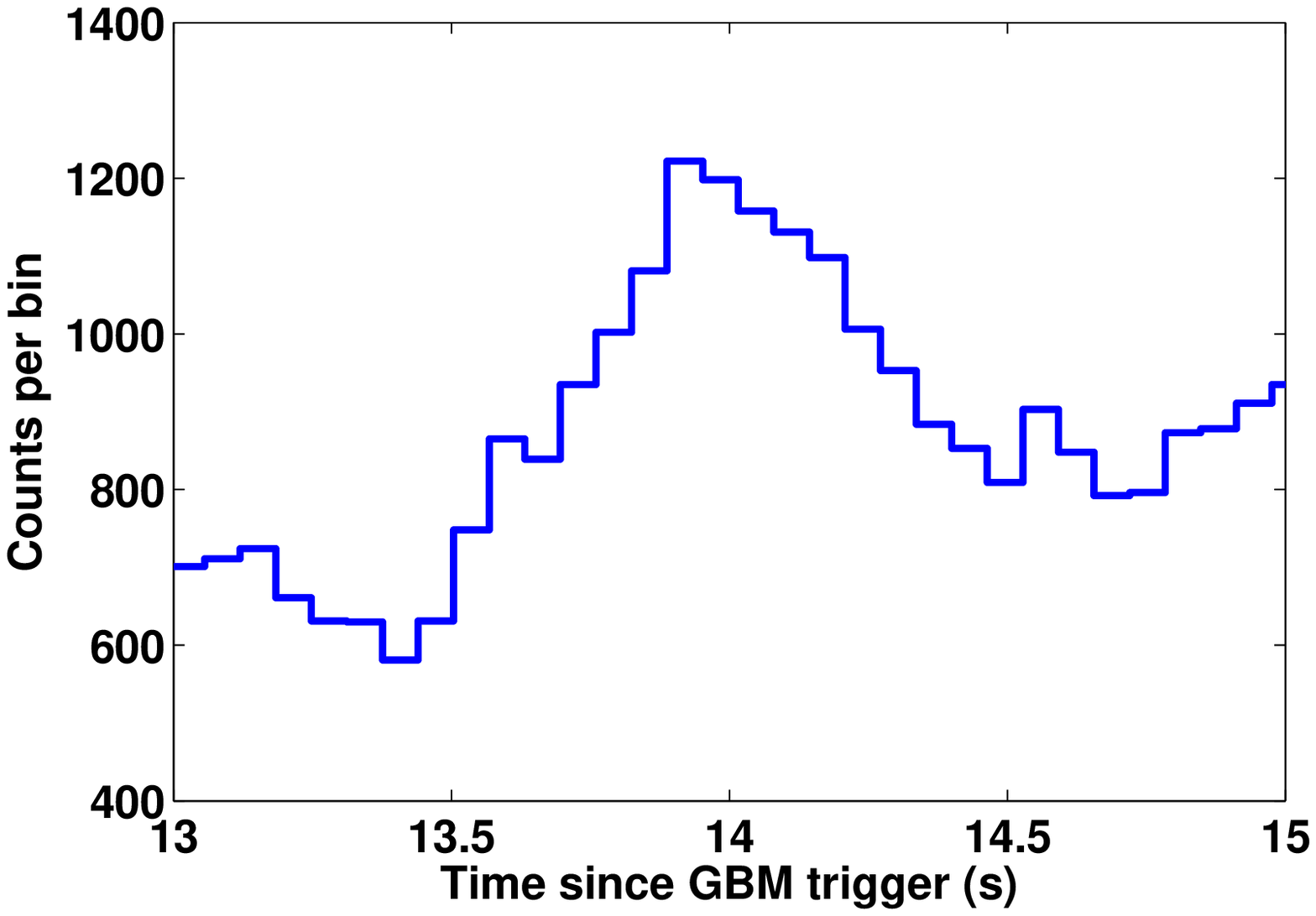}
\end{minipage}
}
\caption{Light curves of the two brightest trigger detectors combined (GBM NaI-n0 and NaI-n3, $8\sim 260~\mathrm{keV}$) for GRB~160509A. In the left panel (a), photon events are binned in 1~second intervals. In the right panel (b), photon events are binned in 0.064~seconds intervals to determine the peak of the main pulse as $T_{\mathrm{peak}}=13.920~\mathrm{s}$.}\label{fig:subfig:1}
\end{figure}

The light curves of the two brightest GBM trigger detectors combined (GBM NaI-n0 and NaI-n3, $8\sim 260~\mathrm{keV}$)~\cite{GBM_data} are shown in Fig.~\ref{fig:subfig:1}. In the left panel Fig.~\ref{fig:1a} where photon events are binned in 1~second intervals, a main pulse lasts about 20 seconds after a first dim short spike near trigger-time. In the right panel Fig.~\ref{fig:1b}, photon events near the peak of the main pulse are binned in 0.064~s intervals to determine the peak time of the main pulse as $T_{\rm peak}=13.920~\rm s$.

GRB~160509A also triggered the Large Area Telescope (LAT)~\cite{LAT} onboard FGST and is located at $\mathrm{(RA, Dec)}=(311.3, 76.1)~\rm (J2000)$ by LAT with a $90~\%$ containment radius of 0.12~degrees (statistical only)~\cite{LAT_localization}. A 51.9~GeV event was observed $T_{\mathrm{arrive}}=76.506~\rm s$ after the GBM trigger. This photon is located at $\mathrm{(RA, Dec)}=(310.3, 76.0)~\mathrm{(J2000)}$ so the directional coincidence of this photon with GRB~160509A is very significant. Several other photons with energy higher than 1~GeV, within the 90 second time window and within a $12^{\circ}$ region of interest (ROI)~\cite{F-LAT_above_10}~are listed in Table~\ref{tab:over_1gev}.

\begin{table}[htbp]
  \centering
  \caption{Photons with energy higher than 1~GeV from  GRB~160509A}
   \begin{tabular*}{0.5\textwidth}{@{\extracolsep\fill}ccc}
    \toprule
   $E_{\rm obs}~/~\rm GeV$ & $t_{\rm arri}~/~\rm s$ & $\rm (RA, Dec)$ \\
    \midrule
    51.9  & 76.506 & $(310.3, 76.0)$ \\
    2.33  & 24.258  & $(313.2, 75.9)$ \\
    1.85  & 87.039  & $(308.3, 73.9)$ \\
    1.52  & 50.570  & $(328.8, 72.5)$ \\
    1.26  & 49.155  & $(311.3,75.8)$ \\
    \bottomrule
    \end{tabular*}%
    \begin{tablenotes}
        \item Photons are selected within the 90 second time window and within a $12^{\circ}$ ROI. $E_{\rm obs}$ is the observed energy while $t_{\rm arri}$ is the arrival time of these photons with respect to the trigger of GBM.
    \end{tablenotes}
  \label{tab:over_1gev}%
\end{table}%

Amelino-Camelia {\it et al.}~\cite{method1, method2} first suggested to use the data of GRBs to test energy dependence of light speed. For photons with energy $E\ll E_{\rm Pl}$, where $E_{\rm Pl} = \sqrt{\hbar c^5 / G}\approx 1.22 \times 10^{19} ~ \rm GeV$ is the Planck scale, leading terms in a Taylor series expansion of the classical dispersion relation are
\begin{equation}\label{eq:1}
  v(E)=c\left[1-s_n\frac{n+1}{2}\left(\frac{pc}{E_{\mathrm{LV,}n}}\right)^n\right],
\end{equation}
where $n=1$ or 2 is usually assumed. $s_n=\pm1$ indicates whether the high energy photon travels slower ($s_n=+1$) or faster ($s_n=-1$) than the low energy photon, and $E_{{\mathrm{LV},}n}$ represents the $n$th-order Lorentz invariance violation~(LV)  scale to be determined by the data. Taking the expansion of the universe into consideration, one has the arrival time lag due to light speed variance between two photons with energies $E_{\rm high}$ and $E_{\rm low}$ respectively as~\cite{newformula,oldformula}
\begin{equation}\label{eq:2}
  \Delta t_{\mathrm{LV}}=s_n\frac{1+n}{2H_0}\frac{E^n_{\mathrm{high}}-E^n_{\mathrm{low}}}{E^n_{\mathrm{LV,}n}}\int_0^z\frac{(1+z')^n\mathrm{d}z'}
{\sqrt{\Omega_{\mathrm{m}}(1+z')^3+\Omega_{\Lambda}}},
\end{equation}
where $z$ is the redshift of the GRB source. $H_0=\mathrm{67.3\pm 1.2 ~km s^{-1} Mpc^{-1} }$ is the Hubble expansion rate and  $[\Omega_m, \Omega_{\Lambda}]=[\mathrm{0.315^{+0.016}_{-0.017}}, \mathrm{0.685^{+0.017}_{-0.016}}]$ are cosmological constants~\cite{pgb}.

Observed time lag consists of both $\Delta t_{\rm LV}$ and the intrinsic time lag $\Delta t_{\rm in}$ at the source of GRBs~\cite{intrinsiclag},
\begin{equation}\label{eq:3}
  \Delta t_{\mathrm{obs}}=t_{\rm high}-t_{\rm low}=\Delta t_{\mathrm{LV}}+(1+z) \Delta t_{\mathrm{in}}.
\end{equation}
The factor $(1+z)$ is due to cosmological expansion. Without further assumptions about the intrinsic properties of GRBs, $\Delta t_{\rm in}$ varies for different photons.

We combine Eqs.~(\ref{eq:2}) and (\ref{eq:3}) as
\begin{equation}\label{eq:4}
  \frac{\Delta t_{\mathrm{obs}}}{1+z}=s_n \frac{K_n}{E^n_{\mathrm{LV,}n}}+\Delta t_{\mathrm{in}},
\end{equation}
where $K_{\mathrm{n}}$ is the Lorentz violation factor
\begin{equation}\label{eq:5}
  K_{n}=\frac{1+n}{2H_0}\frac{E^n_{\mathrm{high}}-E^n_{\mathrm{low}}}{1+z}\int_0^z\frac{(1+z')^n\mathrm{d}z'}
{\sqrt{\Omega_{\mathrm{m}}(1+z')^3+\Omega_{\Lambda}}}.
\end{equation}
We can see that if the energy-dependence of light speed does exist, there would be a linear relation between $\Delta t_{\mathrm{obs}}/(1+z)$ and $K_{n}$. Photons with same intrinsic time lag would fall on an inclined line in the $\Delta t_{\mathrm{obs}}/(1+z)$-$K_{n}$ plot, and we can determine $\Delta t_{\mathrm{in}}$ of them as the intercept of the line with the $Y$ axis. The slope of this line is $s_n /{E^n_{\mathrm{LV,}n}}$, from which we can determine $E_{\mathrm{LV,}n}$.

With known redshift $z=1.17$ and high energy photons, GRB~160509A can help constrain the possible energy-dependence of light speed. Here we follow the works in Refs.~\cite{shaolijing, zhangshu, xhw},  where photons with energies $E>10~\rm GeV$ and within the 90 second time window are adopted as high energy photons described in Eqs.~(\ref{eq:2}) and (\ref{eq:3}), i.e., $E_{\rm high}=51.9~\rm GeV$ and $t_{\rm high}=T_{\rm arrive}=76.506~\rm s$ for GRB~160509A. On the other hand, low energy photons are received constantly during the burst so a unified criterion for $t_{\rm low}$ is required for different GRBs. We also follow Ref.~\cite{xhw} and set $t_{\rm low}$ as the peak time of the main pulse (Fig.~\ref{fig:subfig:1}), which, as a benchmark of a large number of low energy photons, naturally reflects the intrinsic property of GRBs. So $t_{\rm low}=T_{\rm peak}=13.920~\rm s$ for GRB~160509A. Since photons arriving at $t_{\rm low}$ have energies between $8\sim 260~\rm keV$, $E_{\rm low}$ is negligible compared with $E_{\rm high}$. So it is reasonable to set $E_{\rm low}=0$ in Eqs.~(\ref{eq:2}) and (\ref{eq:5}).

\begin{figure}
  \centering
      \includegraphics[width=0.8\textwidth]{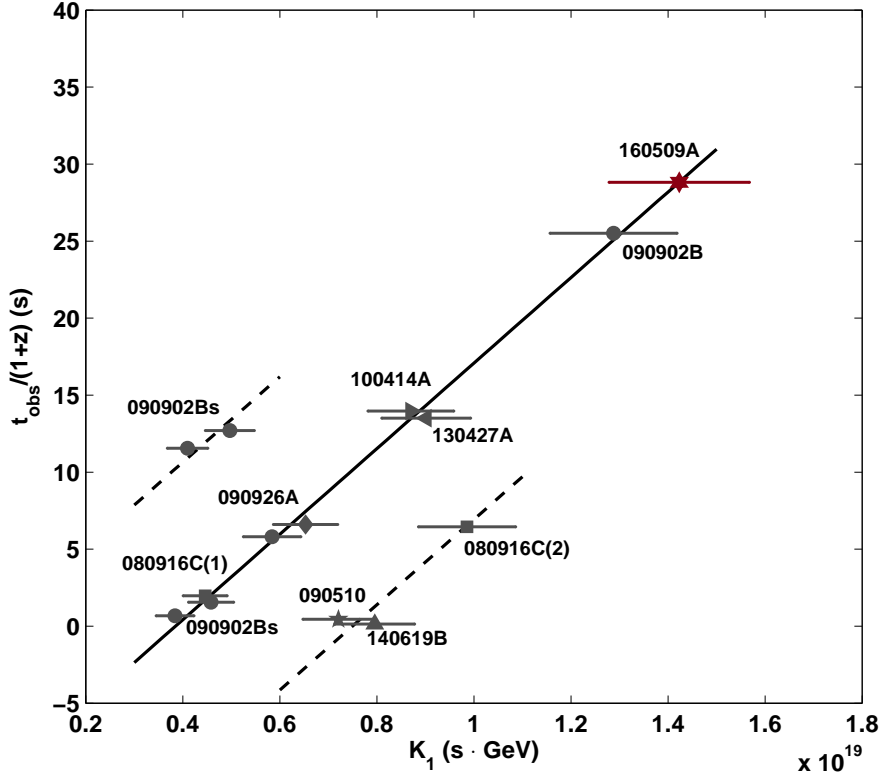}\\
 \caption{The $\Delta t_{\rm obs}/(1+z)$ versus $K_{\rm 1}$ plot for high energy photon events. Events in gray were included in Ref.~\cite{xhw} and the mainline (solid line) was fitted by the eight events on it. The slope was determined as $1/E_{\rm LV,1}=2.78\times 10^{-18} ~\mathrm{GeV^{-1}}$ with $s_{1}=+1$ and the intercept is $\Delta t_{\rm in}=-10.7$~s. GRB~160509A is denoted by the red mark and falls exactly on this mainline. Two dash lines have the same slope as the mainline and their intercepts are $-0.47~\rm s$ and $-20.77~\rm s$ respectively.}\label{fig:2}
\end{figure}

In Refs.~\cite{shaolijing, zhangshu, xhw}, some previous GRBs with high energy photons were adopted and analyzed. In Ref.~\cite{shaolijing}, four events from GRBs~080916C, 090510 (short), 090902B and 090926A were analyzed and three events from long bursts were found to fall on a same line. After that, in Ref.~\cite{data}, all known photon events then with energies greater than 10~GeV, with measured redshifts and within a 90 second window were published (11 events in total, four in Ref.~\cite{shaolijing}, six reconstructed in Pass 8 from aforementioned GRBs and another one from GRBs~100414A). Ref.~\cite{zhangshu}~exhausted all events satisfying above conditions at that time (aforementioned 11 ones and another one from newly detected GRB~130427A,
which was extensively discussed in Ref.~\cite{Amelino}) and revealed a regularity that 5 out of 12 events fall on the mainline. In these two works, the trigger time of GBM was set as the low energy photon arrival time $t_{\rm low}$ in Eq.~\ref{eq:3}. More recently, in Ref.~\cite{xhw}, the peak time of the main pulse of low energy photons was suggested to be $t_{\rm low}$. With this more natural choice of $t_{\rm low}$, a stronger regularity emerges with 8 out of 13 photon events (another event from GRB~140619B was detected and included) falling on the same line in the $\Delta t_{\mathrm{obs}}/(1+z)$-$K_{1}$ plot (see Fig.~1 in Ref.~\cite{xhw}), so a linear form ($n=1$) light speed variation was suggested at a scale of $E_{\rm LV}=(3.60 \pm 0.26) \times 10^{17}~ \rm GeV$.

The corresponding data for the newly detected GRB~160509A energetic photon event are listed in Table~\ref{tab:grbs} in comparison with Table~1 in Ref.~\cite{xhw}. Now we can check the position of this GRB~160509A event to test the prediction from Fig.~1 in Ref.~\cite{xhw}. The results are shown in Fig.~\ref{fig:2}, where the gray marks correspond to previously observed energetic photon events~\cite{xhw} and the red mark is the newly observed GRB~160509A event. It is quite surprising that this GRB~160509A event falls exactly on the mainline (solid line) and the slope of the mainline remains unchanged as $1/E_{\rm LV,1}=2.78\times 10^{-18} ~\mathrm{GeV^{-1}}$ with this additional point. In other words, GRB~160509A strongly supports the conclusion in Ref.~\cite{xhw}.

\begin{table}[t]
\renewcommand\arraystretch{1.5}
\begin{centering}
  \caption{Data of high energy photon event from GRB~160509A}
     \begin{tabular*}{1\textwidth}{@{\extracolsep\fill}cccccccc}
   \hline
   \hline
   GRB         & $z$           & $t_{\rm high}$~{\scriptsize (s)} & $t_{\rm low}$~{\scriptsize (s)} & $E_{\rm obs}$~{\scriptsize (GeV)} & $E_{\rm source}$~{\scriptsize (GeV)} & $\frac{\Delta t_{\rm obs}}{1+z}$~{\scriptsize (s)} & $K_{\rm 1}$~{\scriptsize ($\times 10^{18}~\mathrm{s}~\cdot$~GeV) }\\

    160509A  & $1.17$  & 76.506   & 13.920   & 51.9       &  112.6              & 28.812    & 14.2  \\
   \hline
   \hline
     \end{tabular*}%

   \begin{tablenotes}
        \item Data of GRB~160509A. $t_{\rm high}$ and $t_{\rm low}$ denote the arrival time of the high energy photon event and the peak time of the main pulse of low energy photons  respectively, with the trigger time of GBM as the zero point.
        $E_{\rm obs}$ and $E_{\rm source}$ are the energy measured by Fermi LAT and the intrinsic energy at the source of GRBs, with
       $E_{\rm source}=(1+z)E_{\rm obs}$. $K_{\rm 1}$ is the Lorentz violation factor with a unit of~(s~$\cdot$~GeV) for $n=1$.
    \end{tablenotes}
  \label{tab:grbs}%
  \end{centering}
\end{table}%

In Ref.~\cite{xhw}, it was discussed that there may be other ways to draw parallel lines in Fig.~\ref{fig:2} (see Fig.~3 therein). But with the additional energetic event GRB~160509A, the distribution trend seems to be more significant and one has more confidence to draw the mainline. Furthermore, it is interesting to notice that two events GRB~090902B and GRB~160509A, which play an important role for directing the distribution trend, have identical intrinsic energy $E_{\rm source}=112.6~\rm GeV$, so it is very reasonable to assume that they have identical intrinsic properties at the GRB source, and consequently they have identical intrinsic time lag $\Delta t_{\rm in}$. Therefore one should draw a line across them in the $\Delta t_{\mathrm{obs}}/(1+z)$-$K_{n}$ plot where, as we have mentioned, the intercept of this line with the $Y$ axis indicates the intrinsic time lag $\Delta t_{\rm in}$. It is remarkable that this line in fact coincides with the mainline in Fig.~\ref{fig:2}. This provides a supplementary justification for the mainline as an indication of light speed variation.


Finally, we want to mention that GRB~160509A also supports some other minor conclusions or conjectures in Refs.~\cite{shaolijing, zhangshu, xhw}. The minus sign of $\Delta t_{\rm in}$, as the intercepts of the lines in Fig.~\ref{fig:2}, suggests that some high energy photons are emitted before the intensive pulse of low energy photons at the source of GRBs. Furthermore, the average intrinsic energy $E_{\rm source}=(1+z)E_{\rm obs}$ of photons on the three lines in Fig.~\ref{fig:2}~are $40\pm 4~\rm GeV$, $72\pm28~\rm GeV$ and $96\pm 37~\rm GeV$ respectively (from up to down). Such regularity suggests a
chronological order dependent on intrinsic energy of photons, i.e., the higher intrinsic energy one photon has, the earlier it is emitted, for the three groups of photons cataloged by the three lines in Fig.~\ref{fig:2}. Of course, these conjectures wait for more data to verify.

In conclusion, we analyzed the energetic photon event of a recently detected long GRB~160509A and find that this event strongly supports the prediction for a linear form of light speed variation $v(E)=c(1-E/E_{\mathrm{LV}})$ in cosmological space at a scale of $E_{\rm LV}=3.60\times 10^{17}$~GeV.

\noindent {\bf Acknowledgements}
This work is supported by National Natural Science Foundation of China (Grants No.~11120101004 and No.~11475006) and the National Fund for Fostering Talents of Basic Science (Grant Nos.~J1103205 and J1103206). It is also supported by the Undergraduate Research Fund of Education Foundation of Peking University.


\end{document}